\documentclass[onecolumn,nofootinbib]{revtex4}
\usepackage{graphicx}
\usepackage{amsmath}
\usepackage{amssymb}
\usepackage{float}
\usepackage[all]{xy}
\usepackage{amsfonts}
\usepackage{dcolumn}
\usepackage{bm}
\usepackage{xcolor}
\usepackage{caption}
\usepackage{hyperref}
\usepackage{multirow}
\usepackage[utf8x]{inputenc}
\usepackage{epstopdf}
\usepackage{capt-of}
\usepackage{dcolumn}   
\usepackage{bm}
\usepackage{dsfont}
\usepackage{relsize}
\usepackage{mathtools}

\begin{document}

\title{Growth of matter fluctuations in $f(R,T)$ Gravity }

\author{Snehasish Bhattacharjee \footnote{Email: snehasish.bhattacharjee.666@gmail.com \\ ORCID: https://orcid.org/0000-0002-7350-7043 \\}
  }\

\affiliation{Department of Astronomy, Osmania University, Hyderabad, 500007, India}
\date{\today}

\begin{abstract}
In this work, I present for the first time the analysis concerning the growth of matter fluctuations in the framework of $f(R,T)$ modified gravity where I presume $f(R,T) = R + \lambda T$, where $R$ denote the Ricci scalar, $T$ the trace of energy momentum tensor and $\lambda$ a constant. I first solve the Friedman equations assuming a dust universe ($\omega =0$) for the Hubble parameter $H(z)$ and then employ it in the equation of matter density fluctuations $\delta(z)$ to solve for $\delta(z)$ and the growth rate $f(z)$. Next, I proceed to show the behavior of $f(z)$ and $\delta(z)$ with redshift for some values of $\lambda$ with observational constraints. Finally, following the prescription of \cite{growft41}, I present an analytical expression for the growth index $\gamma$ which is redshift dependent and the expression reduces to $3/5$ for $\lambda=0$, which is the growth index for a dust universe.  
\end{abstract}

\maketitle
  
\section{Introduction}
Cosmological observations favor a spatially flat, isotropic and homogeneous universe with majority of energy-density in the form of some unknown quantities termed dark energy and dark matter \cite{observations}. Regardless of the immeasurable successes in both theoretical and observational aspects of cosmology from the past few decades, the nature of dark energy and dark matter remain unknown and linger on to be the most outstanding problem in theoretical physics. Alternative scenarios have been widely investigated in literature to suffice the late-time acceleration by either modifying the matter-energy content in the so-called scalar field dark energy models which cohere to general relativity, endorsing however the existence of new fields filling the cosmos or by changing the gravitational forces felt by cosmic objects \cite{growft} (see \cite{growft3} for a review).\\
Modified theories of gravity furnish an alternative mathematical treatment by proposing the current acceleration to be due to gravitational effects rather than due to presence of exotic matter-energy sources \cite{growft,growft3}. In such extended theories of gravity, the Ricci Scalar $R$ in the action is replaced by either a well-motivated function of $R$ or by other curvature invariants such as Torsion scalar $\mathcal{T}$ , Gauss-Bonnet scalar $G$, non-metricity $Q$ and trace of energy momentum tensor $T$, which results in modified Friedmann equations with supplementary degrees of freedom which upon tuning fits the observations elegantly. \\   
In this spirit, Harko et al \cite{harko} introduced the $f(R,T)$ gravity which represents a straightforward generalization of $f(R)$ gravity (see \cite{extended} for a review on modified gravity theories). In this theory, the action contain a combined function of $R$ and $T$. $f(R,T)$ gravity has been successful to addressing major cosmological enigmas such as dark matter \cite{in22} dark energy \cite{in21}, massive pulsars \cite{in23}, super-Chandrasekhar white dwarfs \cite{in25}, gravitational waves \cite{in36}, wormholes \cite{in26}, baryogenesis \cite{baryo}, bouncing cosmology \cite{bounce,bounce2}, viscous cosmology \cite{arora}, redshift drift \cite{drift}, inflationary cosmology \cite{inflation} big bang nucleosynthesis \cite{bang} and varying speed of light scenarios \cite{physical}.\\
Since both modified theories of gravity and scalar field dark energy models fit the current observations, one may be interested in distinguishing them. This can be achieved at the perturbation level (see \cite{growft15} for a recent analysis). In particular, the idea of employing the growth index $\gamma$ \cite{growft16} to distinguish modified gravity from the $\Lambda$CDM is well known. One may find numerous studies aimed at deriving the expression of growth index for various cosmological models including DGP \cite{growft21,growft23to25}, scalar field dark energy \cite{growft21,growft17to20,growft22}, Finsler-Randers \cite{growft26}, $f(R)$ gravity \cite{ft27} and $f(\mathcal{T})$ gravity \cite{growft}.\\
Although $f(R,T)$ gravity models have been richly employed to investigate the scalar density perturbations \cite{frt/per}, as far as I know, no analysis concerning the growth rate of clustering have been attempted. In this paper, I shall fill in that gap. The manuscript is organized as follows: In Section \ref{II}, I present an overview of $f(R,T)$ gravity and solve the field equations for $f(R,T)=R+\lambda T$ assuming a dust universe ($\omega =0$). In Section \ref{III} I derive the expressions for $\delta(z)$ and $f(z)$ and show their behavior graphically and also their dependency on the model parameter $\lambda$. In Section \ref{IV} I present an analytical expression for the growth index which is redshift dependent and the expression reduces to $3/5$ for $\lambda=0$, which is the growth index for a dust universe. Finally in Section \ref{V} I summarize the results and conclude the work.

\section{Field Equations in $f(R,T)$ Gravity}\label{II}

The action in $f(R,T)$ is defined as \cite{harko} 
\begin{equation}\label{eq1}
\mathcal{S}=\frac{1}{2\kappa^{2}}\int \sqrt{-g}\left[ f(R,T)+\mathcal{L}_{m}\right] d^{4}x,   
\end{equation}
where $\kappa^{2}=\frac{8 \pi G_{N}}{c^{4}}$. Here  $c$ denotes the speed of light and $G_{N}$ the gravitational constant. Additionally,  $\mathcal{L}_m=-p$ represents matter Lagrangian density with $p$ being the cosmological pressure. In this work, I shall work with natural units and therefore I set $\kappa^{2}=\frac{8 \pi G_{N}}{c^{4}} =1$.\\

Varying the action (\ref{eq1}) with respect to the metric $g_{\mu\nu}$ yields the following field equation

\begin{equation}
\kappa^{2}T_{\mu\nu}-f^{1}_{,T}(R,T)(T_{\mu\nu} + \Xi_{\mu\nu})=\Pi_{\mu\nu}f^{1}_{,R}(R,T)+f^{1}_{,R}(R,T)R_{\mu\nu} -\frac{1}{2}g_{\mu\nu}f(R,T) 
\end{equation}
where $T_{\mu\nu}$ represents the stress-energy-momentum tensor. Note that for a perfect fluid, $T_{\mu\nu}$ can be written as
\begin{equation}
T_{\mu\nu}=- pg_{\mu\nu} + ( \rho + p)u_{\mu}u_{\nu} ,
\end{equation}
where $\rho$ represents energy density, $u_\mu$ the four-velocity and 
\begin{equation}
\Pi_{\mu\nu}= g_{\mu\nu}\square-\nabla_{\mu} \nabla_{\nu}.
\end{equation}
\begin{equation}
\Xi_{ij}\equiv g^{\mu \nu}\frac{\delta T_{\mu \nu}}{\delta g^{ij}},
\end{equation}
where I used the notation $f^{i}_{,X}\equiv \frac{d^{i}f}{d X^{i}}$ .
\subsection{Solutions for $f(R,T) = R + \lambda T$}

For this work I set $f(R,T) = R + \lambda T$, where $\lambda$ is a constant. This is the simplest functional choice of $f(R,T)$ modified gravity as when $\lambda=0$, the field equations correspond to that of GR.  This is also the most extensively studied $f(R,T)$ gravity model and has addressed several shortcomings of the conventional $\Lambda$CDM cosmology (see \cite{harko,in25,bounce,in26,in36} and in references therein). Assuming a flat Friedman-Lem\^aitre-Robertson-Walker space-time with ($-$,$+$,$+$,$+$) metric signature, the modified Friedman equations takes the form

\begin{equation}\label{8_1}
H^{2}= \frac{1}{3}\left[-\frac{\lambda}{2}\left(\omega - 3\right) + \kappa^{2} \right] \rho,
\end{equation} 
\begin{equation}\label{8}
-3H^{2}-2\dot{H} = \left[-\frac{\lambda}{2}\left(1 -3 \omega\right) + \kappa^{2} \omega  \right] \rho,
\end{equation}
where $\omega$ represents the equation of state parameter and overhead dots represent time derivatives.\\

Solving the differential equation \eqref{8} for $p=0$ yields

\begin{equation}\label{9}
H=\frac{\alpha}{t},
\end{equation}
where
\begin{equation}
\alpha = \frac{2 + 3 \lambda}{3 (1+ \lambda)}
\end{equation}
The scale factor $a(t)$ reads 
\begin{equation}
a (t) = a_{0} t^{\alpha}
\end{equation}
Utilizing the relation $a=a_{0}/(1+z)$, where $a_{0}$ is the current scale factor and $z$ the cosmological redshift, I obtain the following time-redshift relation,
\begin{equation}\label{10}
t (z) = \left( \frac{1}{1+z}\right)^{\frac{1}{\alpha}} 
\end{equation}

Substituting \ref{10} in \ref{9}, the Hubble parameter can be expressed in terms of redshift as 
\begin{equation}\label{11}
H (z) = \left[\frac{\left( \frac{1}{1+z}\right)^{\alpha} (2 + 3 \lambda)}{3 (1+\alpha)} \right] 
\end{equation} 

Additionally, the matter energy density $\rho$ reads
\begin{equation}\label{12}
\rho (z) = \left[  \frac{\left( \frac{1}{1+z}\right)^{\frac{-6 (1+\lambda)}{2+ 3 \lambda}} (2+3\lambda) (2+9 \lambda + 8 \lambda^{2})}{3 (1+\alpha)^{2} (1+ 6 \lambda + 8 \lambda^{2})}\right] 
\end{equation}

\section{Growth Rate in $f(R,T)$ Gravity}\label{III}
\begin{figure}[H]
\centering
\includegraphics[width=9 cm]{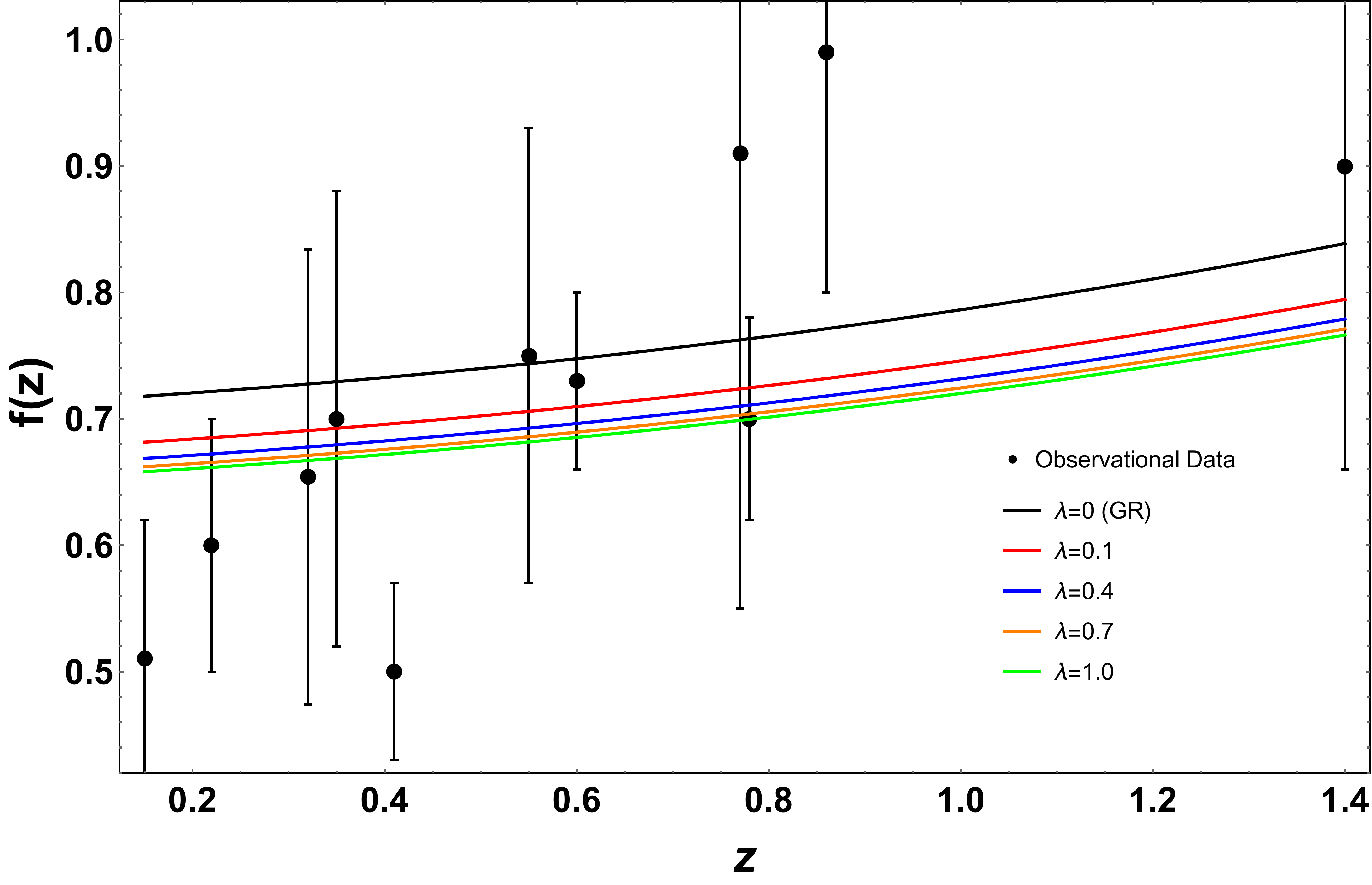}
\caption{Evolution of growth rate $f(z)$ as a function of redshift $z$ for different values of $\lambda$. The profile for $\lambda=0$ corresponds to that of GR. Black dots with error-bars indicate observed values from Table \ref{tab:table}.}
\label{fig1}
\end{figure}

\begin{figure}[H]
\centering
\includegraphics[width=9 cm]{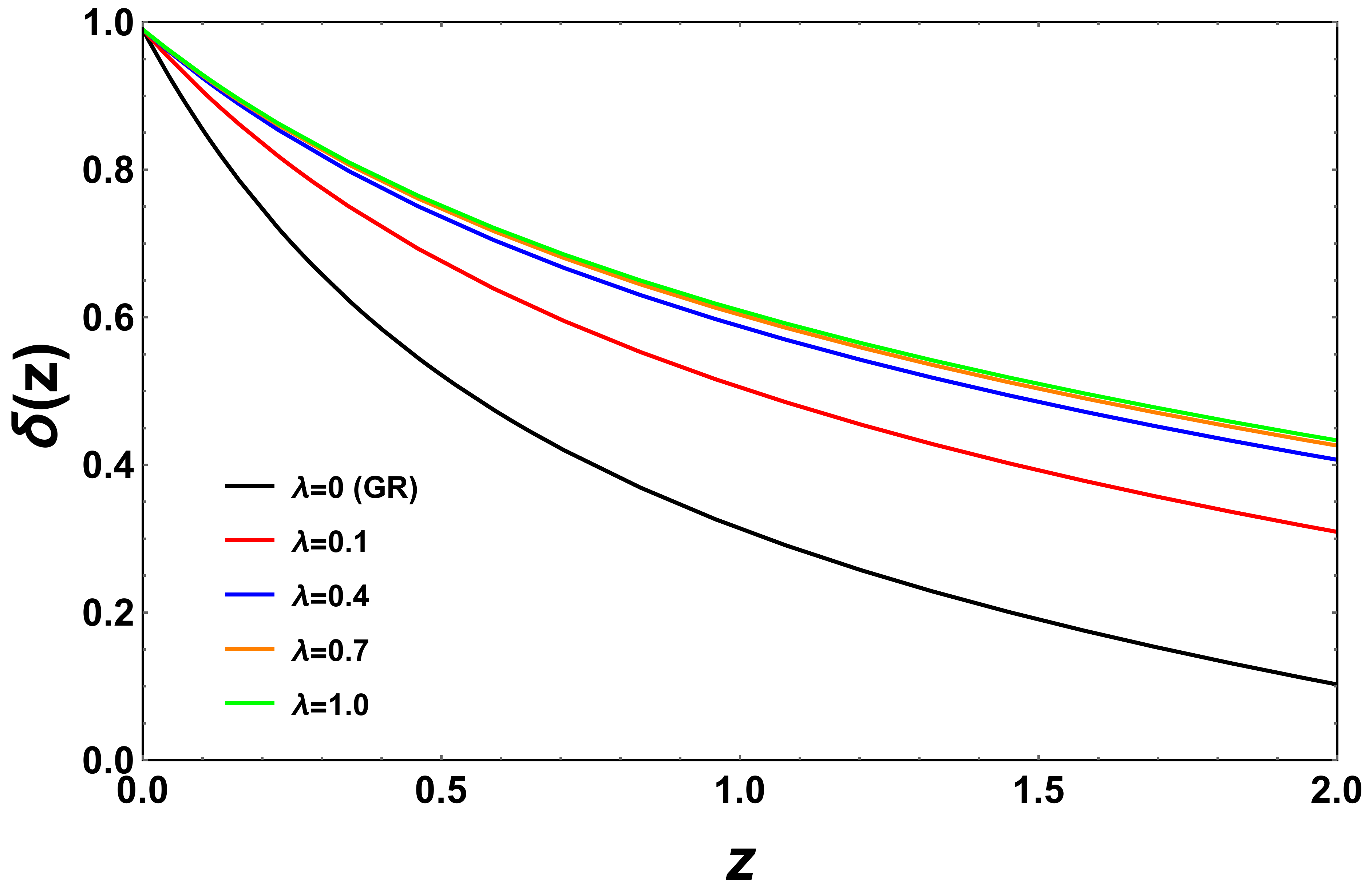}
\caption{Evolution of density perturbations $\delta(z)$ as a function of redshift $z$ for $0\leq\lambda\leq 1$.}
\label{fig2}
\end{figure}

I shall now derive an analytic expression for the matter density fluctuations $\delta(z)$ and growth rate $f(z)$ in the framework of $f(R,T)$ gravity. The second order differential equation describing the evolution of $\delta(z)$ at the sub-horizon scales reads 
\begin{equation}\label{13}
\ddot{\delta} +2 \nu H \dot{\delta} - 4 \pi G_{N} \mu \rho \delta = 0.
\end{equation}
In extended theories of gravity, the quantity $\mu = G_{eff}/ G_{N}$ depends on the scale factor, while for the dark energy models adhered to general relativity, $G_{eff}$ equals Newton's gravitational constant and therefore $\mu=1$ \cite{growft}. $\nu$  is the damping coefficient which is a function of redshift and spatial Fourier frequency \cite{growft41}. Readers are encouraged to see \cite{growft21,growft27,growft38to40,growft41} for further clarifications.\\
Now from Eq. 37 in Ref. \cite{frt/per}, the expression of $G_{eff}$ for $f(R,T) = R + f(T)$ gravity can be written as
\begin{equation}
G_{eff}/G_N=1-f^{1}_{,T}-k^2f^{1}_{,T}/(a^2(1-f^{1}_{,T})\rho). 
\end{equation}
Since I work with natural units, then by setting $G_{N}=1$, the expression of $G_{eff}$ becomes 
\begin{equation}
G_{eff} = 1-f^{1}_{,T}-k^2f^{1}_{,T}/(a^2(1-f^{1}_{,T})\rho).
\end{equation}   
Finally, the parameter $\mu$ for $f(T) = \lambda T$ can be written as
\begin{equation}\label{14}
\mu = G_{eff}/ G_{N} = 1-\lambda -k^2\lambda/(a^2(1-\lambda)\rho).
\end{equation}
It may be noted that for $\lambda = 0$, $\mu = 1$ since $G_{eff} = G_{N}$ as it should be. \\
In most cases, Eq. \ref{13} does not have an analytic solution and can only be approached numerically. Fortunately, an analytic solution does exists for the $f(R,T)$ gravity model employed in this work. Upon substituting the expression of $H(z)$ (Eq. \ref{11}) and $\rho (z)$ (Eq. \ref{12}) in Eq. \ref{13}, the expression of matter perturbation $\delta$ reads 
\begin{multline}\label{15}
\delta (z) =2^{\frac{16 \lambda ^2+3 R+12 \lambda +2}{32 \lambda ^2+24 \lambda +4}} 3^{-\frac{R}{16 \lambda ^2+12 \lambda +2} -\frac{\lambda }{\lambda +1}-\frac{2}{3 (\lambda +1)}} \left(\left(\left(\frac{1}{z+1}\right)^{\frac{3 \lambda }{3 \lambda +2}+\frac{3}{3 \lambda +2}}\right)^{\frac{2}{3 (\lambda +1)}}\right)^{\frac{16 \lambda ^2+R+12 \lambda +2}{32 \lambda ^2+24 \lambda +4}} \left(\left(\frac{1}{z+1}\right)^{\frac{3 \lambda }{3 \lambda +2}+\frac{3}{3 \lambda +2}}\right)^{\frac{3 \lambda +1}{6 \lambda +6}}\\ \times \left((\lambda +1) \left(\frac{1}{z+1}\right)^{\frac{3 \lambda }{3 \lambda +2}+\frac{3}{3 \lambda +2}}\right)^{-\frac{3 \lambda +2}{3 \lambda +3}} \\ \times \left(\frac{\lambda  (\lambda +1)^2  \left(\left(\frac{1}{z+1}\right)^{\frac{3 \lambda }{3 \lambda +2}+\frac{3}{3 \lambda +2}}\right)^{\frac{2}{3 (\lambda +1)}}}{\lambda -1}\right)^{-\frac{R}{32 \lambda ^2+24 \lambda +4}} \left( K_{\frac{R}{16 \lambda ^2+12 \lambda +2}}\left(3 S\right) \times  I_{\frac{R}{16 \lambda ^2+12 \lambda +2}} \left(3 S\right)\right) 
\end{multline} 
where 
\begin{equation}
R = \sqrt{-1728 \lambda ^6-2304 \lambda ^5+1276 \lambda ^4+3576 \lambda ^3+2185 \lambda ^2+546 \lambda +49}
\end{equation}
and 
\begin{equation}
S=\sqrt{\frac{\left(\left(\frac{1}{z+1}\right)^{\frac{3 \lambda }{3 \lambda +2}+\frac{3}{3 \lambda +2}}\right)^{\frac{2}{3 (\lambda +1)}} \lambda  (\lambda +1)^2}{\lambda -1}}
\end{equation}
and $I$ and $K$ represent the modified Bessel functions of the first and second kind respectively. 
The growth rate of clustering $f$ introduced in \cite{growft16} is defined as 
\begin{equation}
f (a) = \frac{d \text{ln} \delta}{d \text{ln} a} = \Omega_{m}^{\gamma} (a)
\end{equation}
which can be re-written in terms of redshift $z$ as 
\begin{equation}\label{16}
f (z) = \frac{- (1+z)}{\delta} \frac{d \delta}{d  z} = \Omega_{m}^{\gamma} (z).
\end{equation}
Where $\Omega_{m}=\frac{8 \pi G \rho_{m}}{3 H_{0}^{2}}$. Substituting Eq. \ref{15} in Eq. \ref{16}, the expression of growth rate becomes 

\begin{multline}\label{17}
f (z) = \frac{\splitfrac{ \splitfrac{3  \lambda  (\lambda +1)^2 I_{\frac{R}{16 \lambda ^2+12 \lambda +2}-1}\left(3 S\right) 
Y+3 \lambda ^3 I_{\frac{R}{16 \lambda ^2+12 \lambda +2}+1}\left(3 S\right) Y +6  \lambda ^2 I_{\frac{R}{16 \lambda ^2+12 \lambda +2}+1}\left(3 \sqrt{\frac{Y \lambda  (\lambda +1)^2}{\lambda -1}}\right) Y}{\splitfrac{ +3 \lambda  I_{\frac{R}{16 \lambda ^2+12 \lambda +2}+1}\left(3 S\right) Y-3  \lambda ^3 K_{\frac{R}{16 \lambda ^2+12 \lambda +2}-1}\left(3 S\right) Y-6  \lambda ^2 K_{\frac{R}{16 \lambda ^2+12 \lambda +2}-1}\left(3 S\right) Y}{\splitfrac{-3  \lambda  K_{\frac{R}{16 \lambda ^2+12 \lambda +2}-1}\left(3 \sqrt{\frac{\left(\left(\frac{1}{z+1}\right)^{\frac{3 (\lambda +1)}{3 \lambda +2}}\right)^{\frac{2}{3 (\lambda +1)}} \lambda  (\lambda +1)^2}{\lambda -1}}\right) Y}{-3  \lambda ^3 K_{\frac{R}{16 \lambda ^2+12 \lambda +2}+1}\left(3 \sqrt{\frac{Y \lambda  (\lambda +1)^2}{\lambda -1}}\right) Y-6  \lambda ^2 K_{\frac{R}{16 \lambda ^2+12 \lambda +2}+1}\left(3 \sqrt{\frac{Y \lambda  (\lambda +1)^2}{\lambda -1}}\right) Y}}}}{
{\splitfrac{\splitfrac{-3  \lambda  K_{\frac{R}{16 \lambda ^2+12 \lambda +2}+1}\left(3 S \right) Y-3  \lambda ^2 \sqrt{\frac{Y \lambda  (\lambda +1)^2}{\lambda -1}} K_{\frac{R}{16 \lambda ^2+12 \lambda +2}}\left(3 S\right)}{ \left(3 \lambda ^2-2 \lambda -1\right) I_{\frac{R}{16 \lambda ^2+12 \lambda +2}}\left(3 S\right) \sqrt{\frac{Y \lambda  (\lambda +1)^2}{\lambda -1}}}}{+ K_{\frac{R}{16 \lambda ^2+12 \lambda +2}}\left(3 S\right) \sqrt{\frac{Y \lambda  (\lambda +1)^2}{\lambda -1}}+2  \lambda  K_{\frac{R}{16 \lambda ^2+12 \lambda +2}}\left(3 S\right) S}}}}{2 (\lambda -1) S (3 \lambda +2) \left( I_{\frac{R}{16 \lambda ^2+12 \lambda +2}}\left(3 S\right)+ K_{\frac{R}{16 \lambda ^2+12 \lambda +2}}\left(3 \sqrt{\frac{Y \lambda  (\lambda +1)^2}{\lambda -1}}\right)\right)}
\end{multline}
where 
\begin{equation}
Y=\left(\left(\frac{1}{z+1}\right)^{\frac{3 (\lambda +1)}{3 \lambda +2}}\right)^{\frac{2}{3 (\lambda +1)}}
\end{equation}
In Fig. \ref{fig1} I show the evolution of Eq. \ref{17} for different values of $\lambda$ and also for $\lambda=0$. the linear growth rate $f(z)$ is a extremely  useful concept in physical cosmology which provides information about the evolution of density perturbations from a nearly smooth state to the highly clumpy state observed at the present epoch. From the figure it is clear that the linear growth rate is sensitive to the model parameter $\lambda$ and it is observed that as $\lambda$ increases, $f(z)$ assumes lower values at all redshift. This implies that the presence of the model parameter $\lambda$ enhances the effect of geometrical dark energy and acts against gravity and represses the growth of formation of structures. Note that for negative $\lambda$, the growth rate becomes imaginary and therefore not permissible. \\
In Fig. \ref{fig2}, I show the evolution of Eq. \ref{15} for $0\leq\lambda\leq 1$ as a function of redshift. I find that, for $\lambda=0$, the density perturbations decrease faster with increasing redshift compared to the cases when $\lambda>0$. Also as $\lambda$ increases, $\delta(z)$ decreases slower. For $\lambda < 0$, $\delta (z)$ increases with redshift which is unphysical. Hence, I find from the present analysis that negative $\lambda$ values do not yield satisfactory results. 
\captionof{table}{Observed growth rate $f(z)$ as a function of redshift $z$.
}\label{tab:table}
\begingroup
\setlength{\tabcolsep}{8pt} 
\renewcommand{\arraystretch}{1.5} 
\begin{center}
\begin{tabular}{ |p{3cm}|p{3cm}|p{3cm}| }
 \hline

  $z$   & $f_{obs} (z)$     & References  \\
    \hline
    0.15 & $0.51\pm0.11$  & \cite{hawkins,verde,linder}    \\
    0.22 & $0.6\pm 0.10$  & \cite{blake}    \\
     0.32 & $0.654\pm 0.18$  & \cite{reyes}    \\
   
    0.35 & $0.7\pm 0.18$  & \cite{tegmark}   \\
    
     0.41 & $0.50\pm 0.07$  & \cite{blake}    \\
     0.55   & $0.75 \pm 0.18$ & \cite{ross}  \\  
    0.60    & $0.73 \pm 0.07$ & \cite{blake}  \\  
    0.77 & $0.91\pm 0.36$  & \cite{guzzo}    \\
     0.78   & $0.7 \pm 0.08$ & \cite{blake}  \\  
     0.86   & $0.400 \pm 0.110$ & \cite{55,104}  \\
    1.4    & $0.90 \pm 0.24$ & \cite{angela}  \\  
    
 \hline
 
\end{tabular}
\end{center}
\vspace{0.5cm}
\endgroup
\section{Growth Index in $f(R,T)$ Gravity}\label{IV}
In this section I shall derive an analytic expression for the growth index $\gamma$ for the $f(R,T)$ gravity model. The growth index is very useful in discriminating various dark energy models from modified gravity theories. It is not clear whether $\gamma$ is a constant or temporally evolving, owing to which, authors have studied the consequence of both a constant growth index and an one which is redshift dependent. However, in this work I shall focus on deriving an expression for a growth index which is redshift dependent. See \cite{growft,growft21,growft23to25,growft17to20,growft22,growft26,ft27}
for other studies related to redshift dependent growth index.\\ 
A general mathematical treatment for the growth index is proposed in \cite{growft41}, according to which the growth index $\gamma_{\infty}$ can be expressed as 
\begin{equation}\label{18}
\gamma_{\infty}=\frac{3 (M_{0}+M_{1})-2 (H_{1}+N_{1})}{2+2 X_{1}+3M_{0}}.
\end{equation} 
Where the relevant quantities are defined as \cite{growft41}
\begin{equation}\label{19}
M_{0}=\mu \bigg|_{\omega=0}, \hspace{0.15in} M_{1} = \frac{d\mu}{d\omega}\bigg|_{\omega=0}
\end{equation}
and \begin{equation}\label{20}
H_{1}=-\frac{X_{1}}{2} = \frac{d (d \text{ln}E / d \text{ln}a)}{d \omega}\bigg|_{\omega=0},\hspace{0.15in}                     N_{1}= \frac{d\nu}{d\omega}\bigg|_{\omega=0},
\end{equation}
where $E(z) = H (z) / H_{0}$ represents the normalized Hubble parameter. For modified gravity theories the quantity $\nu=1$, therefore $N_{1}=0$. \\
Note that in \cite{growft41}, all the cosmological functions are written in terms of the variable $\omega = \text{ln}\Omega_{m}(a)$, implying $\Omega_{m}(a)\rightarrow 1$ when $\omega \rightarrow 0$. Therefore, I can write \cite{growft41,growft}
\begin{equation}
M_{1} = \frac{d\mu}{d\omega}\bigg|_{\omega=0} = \Omega_{m} \frac{d \mu}{d\Omega_{m}} \bigg |_{\Omega_{m}=1}.
\end{equation}
From Eq. \ref{14}, I find for the $f(R,T)$ gravity model, 
\begin{equation}
\mu=-\frac{3 \lambda  (\lambda +1)^2 \left(8 \lambda ^2+6 \lambda +1\right) \left(\frac{1}{z+1}\right)^{\frac{6 \lambda }{3 \lambda +2}+\frac{6}{3 \lambda +2}}}{a^2 (1-\lambda ) (3 \lambda +2) (2 (3 \lambda +1)+\lambda  (8 \lambda +3))}-\lambda +1.
\end{equation}
Now, after some calculations I obtain $\frac{d \mu}{d\Omega_{m}} = 0$ and therefore $M_{1}=0$. \\
Lastly, $H_{1}$ can be written as \cite{growft,growft41}
\begin{equation}
H_{1}=-\frac{X_{1}}{2} = \frac{d (d \text{ln}E / d \text{ln}a)}{d \omega}\bigg|_{\omega=0} = \Omega_{m} \frac{d (d \text{ln}E / d \text{ln}a)}{d \Omega_{m}}\bigg|_{\Omega_{m}=1}.
\end{equation}
For the relevant $f(R,T)$ gravity model used in this work, 
\begin{equation}
d \text{ln}E / d \text{ln}a = -\frac{3 (\lambda +1)}{3 \lambda +2}.
\end{equation}
Since $d \text{ln}E / d \text{ln}a$ is a $constant$, therefore $\Omega_{m} \frac{d (d \text{ln}E / d \text{ln}a)}{d \Omega_{m}}\bigg|_{\Omega_{m}=1} = H_{1}= -\frac{X_{1}}{2} = 0$. Hence I obtain,\\
\{$M_{0}, M_{1}, N_{1}, H_{1}, X_{1}$\} = \{($-\frac{3 \lambda  (\lambda +1)^2 \left(8 \lambda ^2+6 \lambda +1\right) \left(\frac{1}{z+1}\right)^{\frac{6 \lambda }{3 \lambda +2}+\frac{6}{3 \lambda +2}}}{a^2 (1-\lambda ) (3 \lambda +2) (2 (3 \lambda +1)+\lambda  (8 \lambda +3))}-\lambda +1$), $0, 0, 0, 0, 0$\}.\\
Upon substituting all the respective values of the coefficients in Eq. \ref{18}, the expression of growth index in $f(R,T)$ gravity reads
\begin{equation}
\gamma_{\infty}= \frac{3 \left(\frac{3 \lambda  (\lambda +1)^2 \left(8 \lambda ^2+6 \lambda +1\right) \left(\frac{1}{z+1}\right)^{\frac{2}{3 \lambda +2}+2}}{a^2 (\lambda -1) (3 \lambda +2) (\lambda  (8 \lambda +9)+2)}-\lambda +1\right)}{\frac{9 \lambda  (\lambda +1)^2 (2 \lambda +1) (4 \lambda +1) \left(\frac{1}{z+1}\right)^{\frac{2}{3 \lambda +2}+2}}{a^2 (\lambda -1) (3 \lambda +2) (\lambda  (8 \lambda +9)+2)}-3 \lambda +5}.
\end{equation}
For $\lambda=0$, the expression reduces to $\gamma_{\infty} = 3 / 5$, which is the growth index for a dust universe. 
\section{Conclusions}\label{V}
I studied the growth of matter fluctuations in the framework of $f(R,T)$ modified gravity for the simplest functional choice  $f(R,T) = R + \lambda T$, where $R$ denote the Ricci scalar, $T$ the trace of energy momentum tensor and $\lambda$ a constant. This is by far the most widely studied $f(R,T)$ gravity model in the literature \cite{harko,in25,bounce,in26,in36}. However, to the best of my knowledge, no studies related to matter density fluctuations have been attempted in $f(R,T)$ gravity thus far.\\
To carry out the analysis, I first solve the field equations assuming a dust universe ($\omega =0$) for the Hubble parameter $H(z)$ and employ it to solve the equation of matter density fluctuations and finally for the growth rate $f(z)$ and show their behavior with redshift for some values of $\lambda$ with observational constraints on $f(z)$. I found that positive $\lambda$ yields remarkable values of $f(z)$ which suits perfectly with observations, while negative $\lambda$ makes $f(z)$ imaginary and therefore should be refrained from using in future studies. I find that as $\lambda$ increases, $f(z)$ assumes lower values at each redshift. \\
Furthermore, I find that, for $\lambda=0$, the density perturbations decrease faster with increasing redshift compared to the cases when $\lambda>0$. Also as $\lambda$ increases, $\delta(z)$ decreases slower. For $\lambda < 0$, $\delta (z)$ increases with redshift which is unphysical.\\
Additionally, following the prescription of \cite{growft41}, I also present in this work an analytical expression for the growth index which is redshift dependent and the expression reduces to $3/5$ for $\lambda=0$, which is the growth index for a dust universe. 

\section*{Acknowledgments}
I thank Tiberiu Harko for helpful suggestions. I thank the anonymous reviewer for useful criticisms and encouraging comments that helped me to improve the work significantly. I acknowledge DST, New-Delhi, Government of India for the provisional INSPIRE fellowship selection [DST/INSPIRE/03/2019/003141].

\end{document}